\documentclass[traditabstract]{aa} 
\usepackage{graphicx,natbib,longtable,txfonts,lscape}
\usepackage[switch]{lineno}

\newcommand{\kms}{$\rm{\,km \,s}^{-1}$}

\newcommand{\ergscm}{\>{\rm erg}\,{\rm s}^{-1}\,{\rm cm}^{-2}}

\newcommand{\lya}{Ly$\alpha\,$}

\usepackage{color}

\begin{document}
\title{The quest for high-redshift radio galaxies. \\ I. A pilot
  spectroscopic study} \author{Alessandro Capetti\inst{1} \and
  Barbara Balmaverde\inst{1} \and Miguel Coloma Puga\inst{2,1} \and
  Bruno Vizzone\inst{3} \and Ana Jimenez-Gallardo\inst{4} \and Abigail
  Garc\'ia-P\'erez\inst{2} \and Giacomo Venturi\inst{5} }

\institute{INAF - Osservatorio Astrofisico di Torino, Strada
  Osservatorio 20, I-10025 Pino Torinese, Italy, \and Dipartimento di
     Fisica, Universit\`a degli Studi di Torino, via Pietro Giuria 1,
     10125 Torino, Italy \and
School of Physics, Georgia Institute of Technology, 837 State Street
Atlanta, USA, Georgia 30332-0430
  \and European Southern Observatory, Alonso de Córdova 3107,
  Vitacura, Región Metropolitana, Chile
\and Scuola Normale Superiore, Piazza dei Cavalieri 7, I-56126 Pisa, Italy}

\date{}

\abstract{The population of high-redshift radio galaxies (HzRGs) is
  still poorly studied because only a few of these objects
  are currently known. We here present the results of a pilot project of
  spectroscopic identification of HzRG candidates. The candidates are
  selected by combining low-frequency radio and optical surveys that
  cover a total of $\sim 2,000$ deg$^2$ using the dropout
  technique, that is, the presence of a redshifted Lyman break in their
  photometric data.  We focused on 39 $g$-dropout sources, which is about one
-third of the selected sources, that are expected to be at $3.0 < z < 4.5$.

  We considered single and double radio sources separately and searched for $g$-dropout sources at the location of the midpoint of the radio structure for the latter. The host galaxy is expected
  to be located there. We confirm only one out of 29 candidate HzRG associated with an
  extended radio source. For the compact radio
  sources, we instead reach a success rate of 30\% by confirming 3 out of 10
  HzRG targets.

  The four newly discovered HzRGs show a wide range of spectral radio
  slopes. This supports the idea that not all HzRGs are ultrasteep radio
  sources (USSs). The criterion for USSs is most commonly used to find
  HzRGs, but this method only selects a subpopulation. We discuss
  various contamination sources for the objects that are selected with the
  Lyman-break method and conclude that they are likely mainly HzRGs,
  but with a \lya\ line that is underluminous with respect to expectations. }

\maketitle

\section{Introduction}
\label{intro}

Powerful radio-loud active galactic nuclei (RLAGN) represent the most
extreme manifestation of accretion onto a supermassive black hole, and they
play a crucial role in galaxy evolution. Their host galaxies are giant
ellipticals that host the most massive supermassive black holes
in the Universe \citep{best05a,chiaberge11}. The nuclear emission and relativistic jets of RLAGN influence the star formation history in
their host and the energy balance of the intracluster medium (ICM;
e.g., \citealt{voit15,fabian12}). This AGN feedback is required in
numerical simulations to match their predictions with the observations
of the galaxy luminosity functions (e.g., \citealt{croton06}).

The RLAGN at high redshift provide unique diagnostics for conditions in
the early Universe. They represent beacons for finding
distant massive galaxies and protoclusters, for example, which enables us to explore
their properties and space density evolution. It is becoming
increasingly clear that most high-z RLAGN are
obscured. \citet{volonteri11} first noted that the number of RL
quasars is smaller than expected based on the number of blazars found
at high redshift. To solve this difference, \citet{ghisellini16} proposed
that a large fraction of high-redshift RLAGN may be obscured. This
scenario has recently been strengthened by the study of \citet{capetti24},
who estimate that as many as 90\% of RLAGN at z$>$3.5 are obscured in
the optical and UV bands.  Several studies (see, e.g.,
\citealt{merloni14,vijarnwannaluk22}) suggested that most high-redshift
AGN, and not just the RL subclass, are obscured.

These results imply that the majority of the high-z RLAGN appear
as high-redshift radio galaxies (HzRGs), in which the active nucleus
is hidden by circumnuclear absorbing material. While large samples of
radio galaxies (RGs) with well-defined selection criteria are
available at low redshift ($z \lesssim 1$) and their properties were studied in great detail (see, e.g., \citealt{tadhunter16}), our
knowledge of HzRGs is extremely limited. Only 19 RGs at $z>3.5$ were
listed by \citet{miley08} in their review (and just six at z$>$
4), and only five spectroscopically confirmed HzRGs were subsequently
added to this list (e.g., \citealt{jarvis09,saxena18,yamashita20}).

Several questions concerning these powerful RLAGN await an answer. For
example, while their comoving space density increases dramatically, by
a factor of 100 - 1000 from the local Universe to $z \sim$ 2
\citep{willott01}, there are still huge uncertainties in their
evolution at higher redshifts (e.g., \citealt{massardi10}). Another
issue that can be addressed by studying HzRGs are the properties
of their host galaxies. For example, we can test whether they represent the high end of
the galaxy luminosity function at high
redshift as well, as is observed in nearby RLAGN (as suggested by
  \citet{rocca04}). This test cannot be performed on RL quasars because the nuclear light strongly contaminates them.

The aim of this project is to explore the best strategy for building a
statistically sound sample of $z \gtrsim 3.5$ HzRGs that enables
us to explore their properties based on a robust statistical
basis. Most of the known HzRGs were found based on
spectroscopy of the optical counterparts of ultrasteep sources (USS)
in the radio band (e.g.,
\citealt{rottgering97,saxena18}).\footnote{The authors variously defined
  USS with a threshold of $\alpha < -1.3$ \citep{saxena18}
  or $\alpha < -1$ \citep{rottgering97}.}  This method is motivated by
the phenomenological trend of the apparent increase in the radio
spectral index with redshift \citep{miley08}. However, some of the
newly discovered HzRGs differ from this trend, which indicates that
the search for HzRGs among the USS might lead to the selection of a
biased subsample of this population.

We follow a different approach that was widely used to find high-z
sources. This approach is based on the Lyman-break technique, that is, on the drop in flux
of a given source that is caused by the absorption of neutral hydrogen along our
line of sight (see, e.g., \citealt{steidel95}). The drop occurs at
rest wavelengths shorter than that of the \lya\ and for
sufficiently high redshifts falls into different optical or near-infrared
(NIR) bands. For example, for objects at z$\sim 3.0-4.5$, the break is
located at $\lambda \sim 5000-6600$ \AA, which causes them to appear as
$g$-band dropouts.

The paper is organized as follows: In Sect. 2 we describe the method
we used to select the HzRG candidates. The results from the observations
of a candidate subsample are reported in Sect. 3. In
Sect. 4 we explore possible explanations for the relatively low
success rate of the spectroscopic identifications. The results are
discussed in Sect. 5, and we summarize and conclude in
Sect. 6.  We adopt the following set of cosmological parameters: $H_0$
= 69.7 \kms Mpc$^{-1}$ and $\Omega_m$=0.286 \citep{bennett14}.

\section{Selection of the HzRG candidates}
\label{sample}
The HzRG candidates were selected by taking advantage of recent
large area radio and optical surveys. On the optical side, we
concentrated on two surveys: the Hyper Suprime-Cam Subaru Strategic
Program survey (HSC-SSP; \citealt{aihara18}) and the fourth data release
(DR4) of the Kilo-Degree Survey (KiDS; \citealt{kuijken19}). The wide
HSC-SSP covers 1400 deg$^2$ in five broad bands ($g$, $r$, $i$, $z$,
and $y$) with a 5$\sigma$ magnitude limit in the $r$ and $g$ bands of
$r \sim 26.1$ and $g \sim$ 26.5 in the AB system,
respectively. Two fields of the wide HSC-SSP (the Spring and Fall
equatorial fields, located in the regions $09^{\rm h} 00^{\rm m} \leq$
RA $\leq 15^{\rm h} 30^{\rm m}, -2^\circ \leq$ Dec $\leq +5^\circ$ and
$22^{\rm h} 00^{\rm m} \leq $RA$ \leq 02^{\rm h} 40^{\rm m}$ ,
$-1^\circ \leq $ Dec $\leq 7^\circ$, respectively) cover a total of
1310 deg$^2$. After completion, the KiDS \citep{dejong13} will cover
1,500 deg$^2$ in the $u$,$g$,$r$, and $i$ bands with a depth at
5$\sigma$ of $r$=25.2 and $g$=25.4. The two main fields of KiDS,
KiDS-N and KiDS-S, are two wide strips centered at RA $\sim 13^{\rm
  h}$ and DEC$\sim 0$ and RA $\sim 1^{\rm h}$ and DEC$\sim -30^\circ$,
respectively. The latest data release (DR4) covers 993 deg$^2$. KiDS-N and the
Spring equatorial fields of the HSC-SSP overlap significantly.

The spectroscopy program was divided into two runs, with slightly
different criteria for the candidate selection. For the first run of
optical spectroscopic observations, our search started with
observations obtained with the Murchison Widefield Array (MWA;
\citealt{tingay13}) for the whole sky area with $\delta<30^\circ$ in
the frequency range 72-231 MHz, to produce the Galactic and
Extra-Galactic All-Sky MWA Survey (GLEAM, \citealt{wayth15}). The
selection at low frequency was motivated by the need to sample the
rest-frame gigahertz radio emission in high-redshift sources and by
the suggestion that HzRGs are mostly found among USSs, which are
therefore best visible at low frequency.  \citet{hurley17} produced
the extragalactic GLEAM catalog (EGC) by imposing a Galactic latitude
$|b| > 10^\circ$. The EGC consists of 307,455 sources that were
selected from images at 200 MHz with a spatial resolution of
$\sim2^\prime$, and it reaches a completeness of $\sim 55\%$ at $\sim$
50 mJy.

We then selected the EGC sources with F $>$0.3 Jy at 76 MHz within
the HSC-SSP and KiDS area. This flux density threshold corresponds
to a luminosity at the rest-frame frequency of $\sim
370$ MHz of $\sim 5 \times 10^{28}$ W Hz$^{-1}$ for an RG at z=4, which is a typical value for
known HzRGs with z$\sim 3 -5$ \citep{miley08} and for the most luminous
radio sources in the third Cambridge catalog at z$\lesssim 1$. The
completeness of the EGC above this flux density threshold is $>$95\%.

For the second run, we slightly modified the radio selection. We still
considered a low-frequency survey, the Tata Institute of Fundamental
Research Giant Metrewave Radio Telescope (\citealt{swarup91,intema17}), called
TGSS (Tata Institute of Fundamental Research GMRT Sky Survey), which was
performed at 150 MHz but has a better spatial resolution ($\sim
25\arcsec$) and a lower flux density limit (17.5 mJy) than
the EGC. We selected all TGSS sources above a threshold of 50 mJy, which is fainter by a
factor of $\sim$ 3 than those selected in the EGC (assuming a
spectral index of $-0.8$ between 76 and 150 MHz).

The spatial resolution of the GLEAM and TGSS images is
insufficient to locate the host galaxy of the radio source. We then
used the catalog of radio sources built from the observations obtained
for the Karl G. Jansky Very Large Array Sky Survey (VLASS;
\citealt{lacy20}). The VLASS produced multi-epoch images at 3 GHz of
the sky area with DEC$>-40^\circ$ with a resolution of 2\farcs5 and a
rms in the coadded data of 70 $\mu$Jy/beam. We extracted all sources within a distance of 30\arcsec\ from the GLEAM or
TGSS positions from the catalog.

\begin{figure}
  \center
  \includegraphics[width=0.4\textwidth]{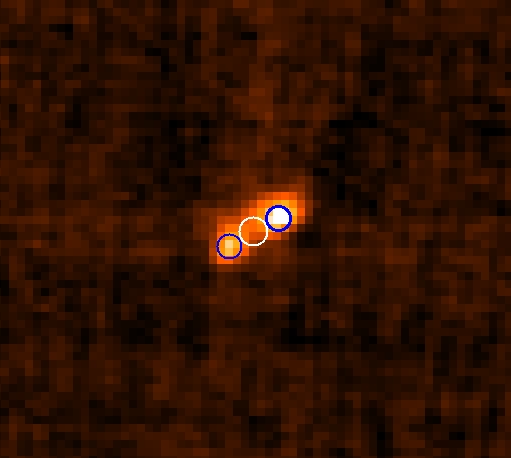}
    \caption{Example of a radio source (NVSS~J221708-325449)
      associated with multiple VLASS radio components. The blue
      circles represent the search area for optical counterparts at
      the position of the two radio components, and the white circle
      is centered at their geometric midpoint. The field of view
        of the image is $1^\prime \times 1^\prime$.}
    \label{search}
\end{figure}

The search for the optical counterparts was performed by adopting
different approaches depending on the properties of each radio
source. When a single compact VLASS radio component
was associated with a given GLEAM or TGSS source, we used a search radius of
1\farcs5. This value was suggested by the uncertainties of the relative
astrometry of radio and optical surveys. When the single component was
extended, we used a value for the search radius that was twice the measured
full width at half maximum, with a minimum value of 1\farcs5.

When we found more than one component, the components might represent the two lobes of a double radio galaxy. This
suggestion is supported by the large fraction of double-lobed
sources among the known HzRGs. In the list of \citet{miley08}, almost
half of the HzRGs with $z>3$ are double radio sources. We
searched for the host galaxy not only at the location of all the
individual components in these cases, but also close to their geometric midpoint.
Several studies (e.g.,\citet{barthel88} and \citet{cordun23})
  showed that high-redshift radio sources are strongly distorted, for instance,
  by asymmetry and bending. We therefore did not only search for the
  host galaxy at the midpoint and along the connecting line
  between the radio lobes, but adopted a search radius of one-sixth
of the distance between the radio components. This value ensured that we
included the host galaxy in the search area even for asymmetric
sources, for which the asymmetry ratio can reach 1:3 between the two
arms. The minimum search radius was again set to 1\farcs5. When we
found multiple VLASS components, we adopted this strategy for all
pairs of radio sources. In Fig. \ref{search} we show an example of the
search areas for a radio source that is associated with multiple VLASS
components.
 
We dropped the sources from the list of candidates with an available
spectroscopic redshift. This occurred in a few
cases, and they all lie at $z<1$. We then visually inspected the radio
and optical images of all HzRG candidates. We discarded sources showing
a complex morphology in either of the two bands. Finally, we excluded
all sources with an $r$-band magnitude $<20$, which are likely to be radio-loud quasars.

After we defined the search areas associated with each radio source, we
queried the HSC-SSP and KiDS catalogs in a search for host galaxies at
high redshift. At this stage, two different approaches are possible. One approach is
based on color-color diagrams (see, e.g., \citealt{ono18,pouliasis22}), and the other on
estimates of photometric redshifts. As shown in the next section, the
photometric redshifts are not sufficiently accurate for a
selection of HzRG candidates.

\subsection{Photometric redshifts}

A widely used method for selecting high-redshift sources relies on
estimates of the photometric redshifts (see, e.g.,
\citealt{koo85}). Photometric redshifts are based on all the available
photometric measurements and are not limited to three bands, as in the
color-color technique described in the next subsection. They might therefore provide us with a more accurate
selection of HzRG candidates in principle.

\begin{figure}
    \includegraphics[width=0.5\textwidth]{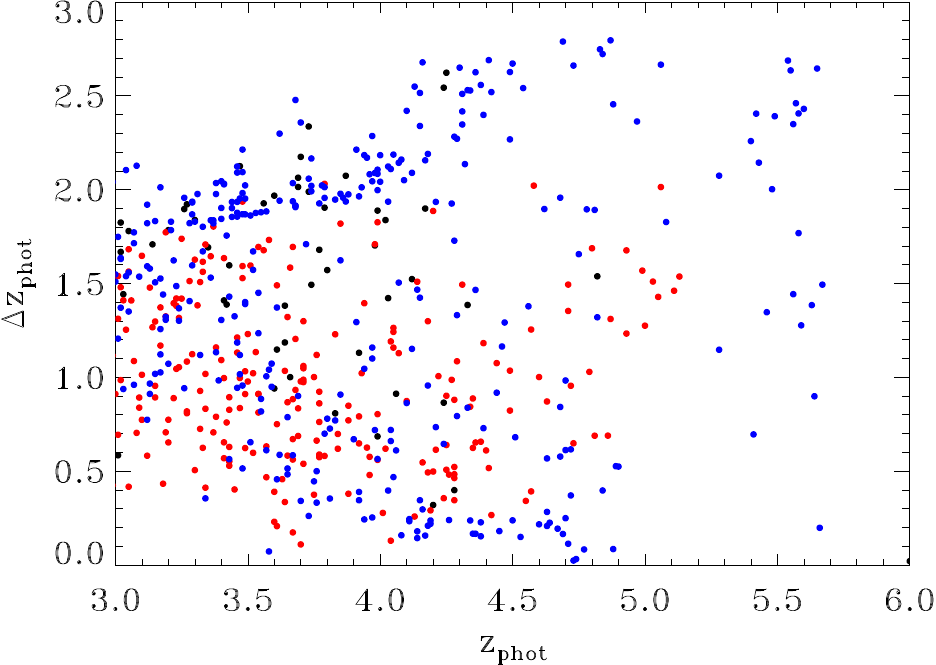}
    \caption{Estimates of the photometric redshifts for the optical
      counterparts of radio sources in the HSC-SSP area compared to
      the errors in this measurement. The symbol colors refer
        to the three different methods: black (MIzuki), red (DemP),
       and blue (dNNz).}
    \label{zphot}
\end{figure}

In Fig. \ref{zphot} we report the photo-z estimates for all the
optical counterparts of radio sources, limited to those in
the HSC-SSP area. The survey catalog lists these estimates and
their errors \citep{tanaka18}. The reported values were obtained
  with three different methods: 1) MIzuki \citep{tanaka15}, 2) the direct
  empirical photometric code (DemP; \citealt{hsieh14}), and 3) deep neural
  network photo-z (dNNz; \citealt{nishuzawa20}).  This figure
indicates that at high redshifts, the $z_{\rm phot}$ errors are often
large. Furthermore, the estimates derived from the various methods
often differ by a very large amount, indicating that they are highly
model dependent. These uncertainties are mostly due to the quite large
errors in the photometric measurements of these faint sources, more
than half of which have an $r$ -band magnitude larger than
23. Therefore, the selection of HzRG candidates based on photometric
redshifts does not appear to be a robust approach.

\subsection{$g$-dropouts in color-color diagrams}

We selected $g$-band dropouts that met the criteria proposed by
\citet{ono18} and \citet{pouliasis22}, that is,
$$g-r > 1.0, \, r-i< 1.0 {~~~\rm and~~~} g-r> 1.5 \times (r-i) + 0.8$$,
which are expected
to be at $3.0 \lesssim z \lesssim 4.5 $. When a source was
undetected in the $g$ band, we adopted the depth at 5$\sigma$ as a lower
limit to its magnitude. 

\begin{figure}
    \includegraphics[width=0.5\textwidth]{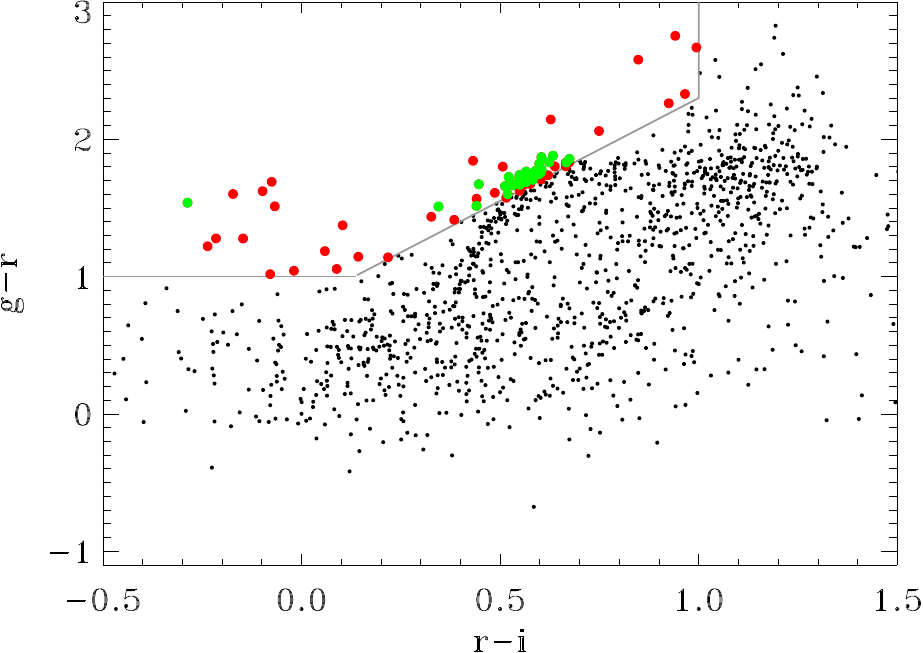}
    \caption{Example of color-color ($r-i$ vs. $g-r$) plot of the
      optical sources in the KiDS area that are counterparts of TGSS
      radio sources. We indicate the sources that meet the criteria
      for a $g$-band dropout (red) and that meet the more
      conservative criterion (green), requiring that the whole ellipse defined
      by the 1$\sigma$ errors falls into the $g$-dropout area.}
    \label{gri}
\end{figure}

In Fig. \ref{gri} we show the $g-r$ versus $r-i$ color-color plot of
all optical sources located within the search areas limited to the optical counterparts of the TGSS sources in the KiDS
area as an example. The sources that meet the g-dropout criteria are marked with a
red circle. For the second run of spectroscopic observations, we
adopted more conservative criteria and required that the whole ellipse
defined by the 1$\sigma$ errors falls into the $g$-dropout area. We
marked these sources with green circles.

\section{Spectroscopy program}

\begin{figure}
  \center
    \includegraphics[width=0.48\textwidth]{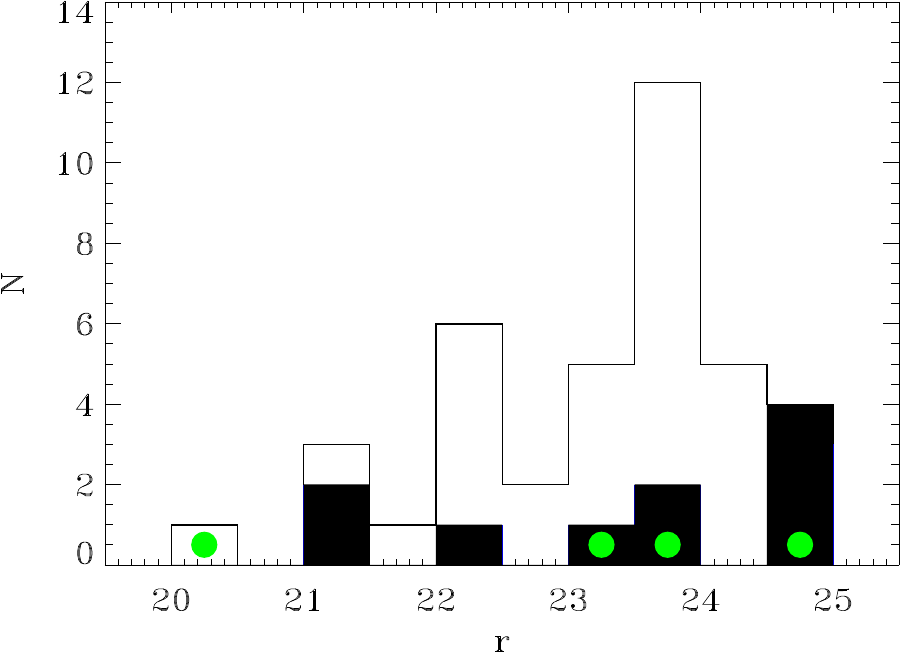}
    \caption{Distribution of the $r$ -band magnitude of the 39 selected
      $g$-band dropout sources observed at the NTT. The filled
      portion of the histogram shows the optical targets
      associated with a compact radio sources, and the four green circles
      represent the confirmed HzRGs.}
    \label{histograms}
\end{figure}

For the first observing run, we selected 30 sources for the
spectroscopy program that were needed to assess the nature of the candidate
HzRGs that were selected based on the color-color diagram. We considered the
sources that are best observable at the ESO New Technology Telescope (NTT) at
the date of the observations (program ID 111.24L3.001; PI: Alessandro
Capetti) to be observed in groups of 10 for each of the three
awarded nights, from August 10 to August 12,
2023. During the first night, the seeing was in the range 2\arcsec\ -
3\arcsec\,, which prevented us from obtaining useful data, so that only 20
sources were observed. The observations were carried out with
the ESO Faint Object Spectrograph and Camera (EFOSC2) spectrograph
through a long slit with a width of 1\arcsec\. For the first run, we used grism 11,
which covers the spectral range 3380-7520 \AA  and provides a dispersion
of 2.04 \AA\ per pixel. In the second run (program ID 114.26ZD.001;
PI: Alessandro Capetti), 3.7 nights from October 28 to
November 1, 2024, we observed 19 HzRG candidates with grism
13 to extend the spectral range to 3685-9315 \AA, with a dispersion
of 2.77 \AA\ per pixel and through a long slit with a width of 1\arcsec\. In most
cases, the target was too faint to be visible in short acquisition
images, and the targets were located within the slit after a blind offset of a
few arcseconds from a brighter nearby source. Two exposures of
$\sim$1,300 seconds each were obtained. These 39 sources represent
about one-third of all HzRG candidates we selected. The distribution of the
$r$ -band magnitude of the observed targets is reported in
Fig. \ref{histograms}. The median value of these sources is $r=23.5$.

\begin{figure*}
  \center
    \includegraphics[width=0.24\textwidth]{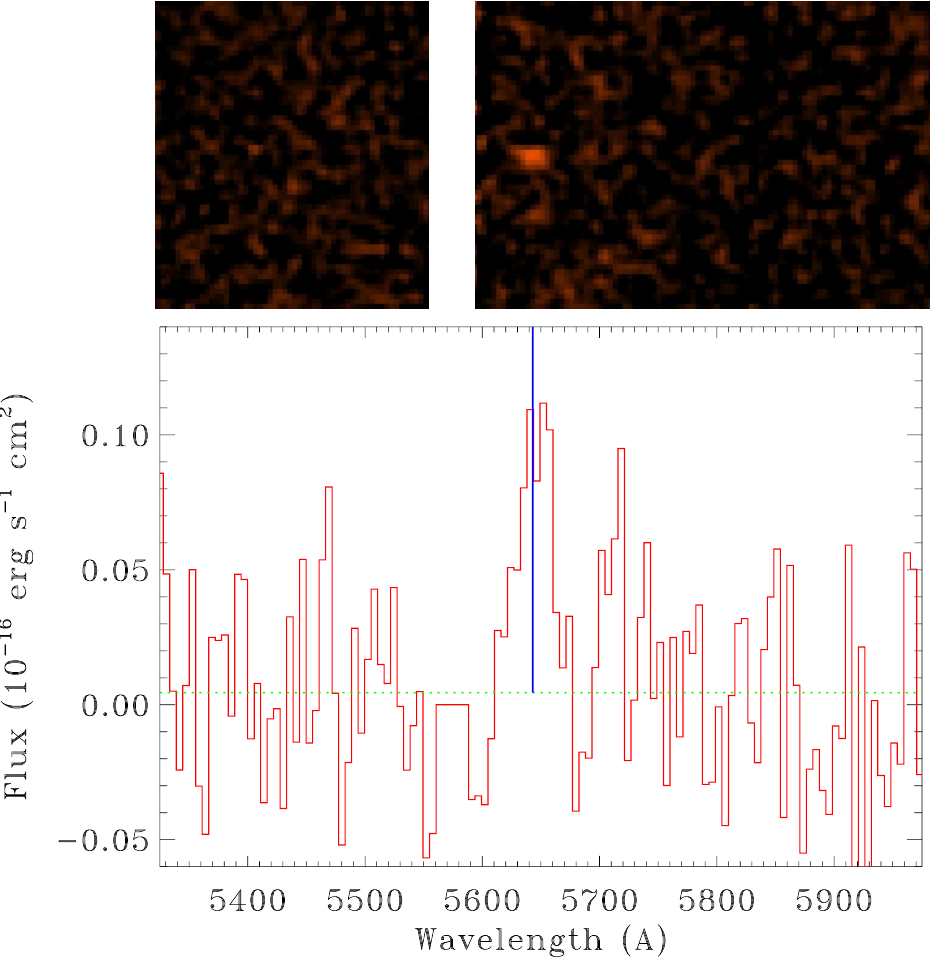}
    \includegraphics[width=0.24\textwidth]{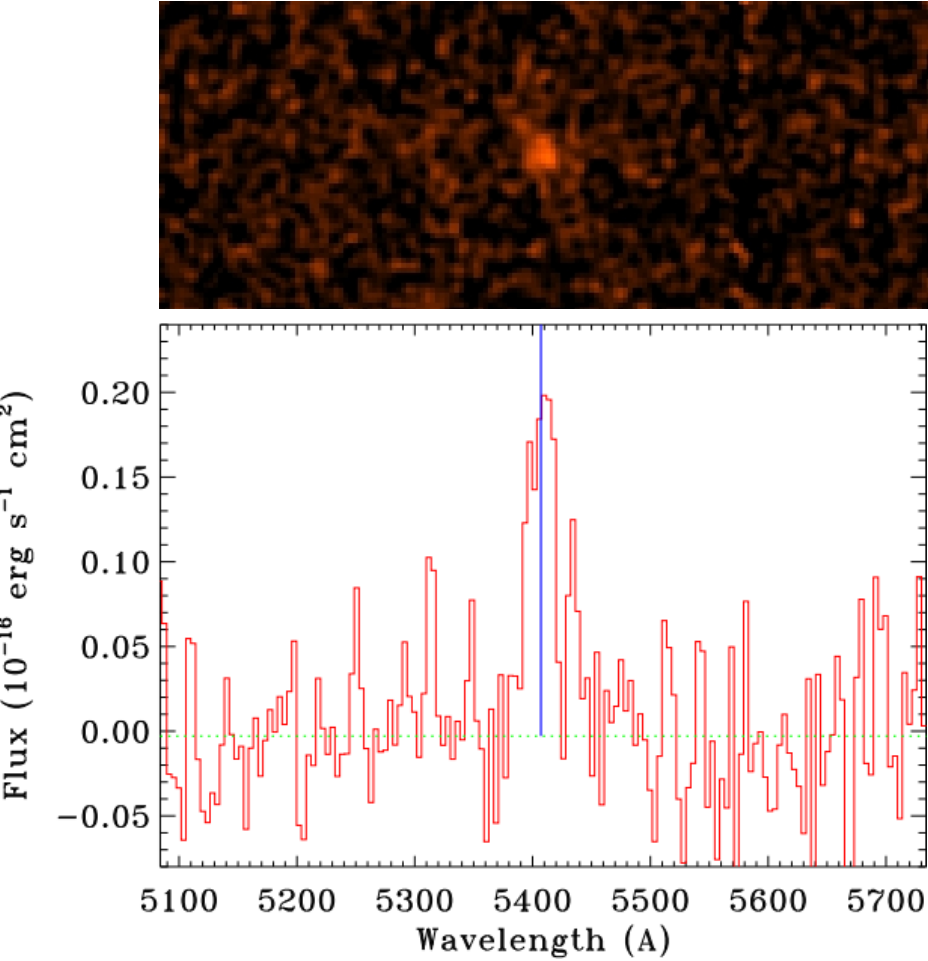}
    \includegraphics[width=0.24\textwidth]{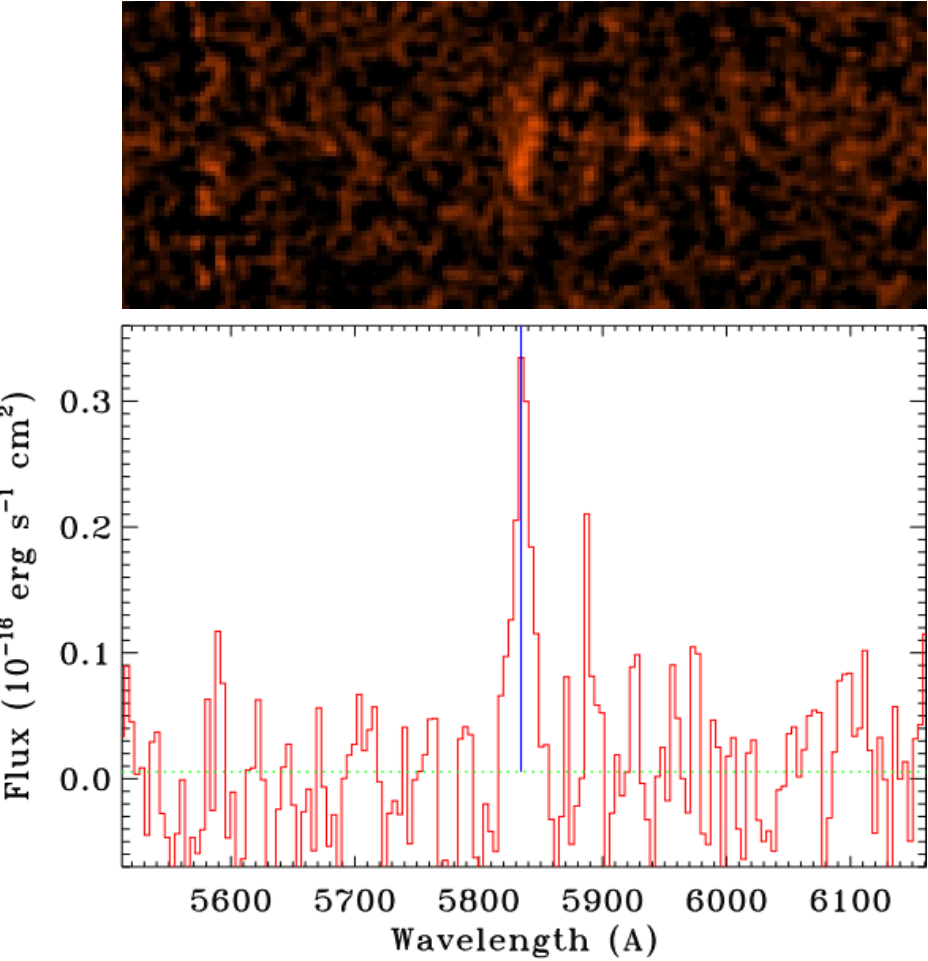}
    \includegraphics[width=0.24\textwidth]{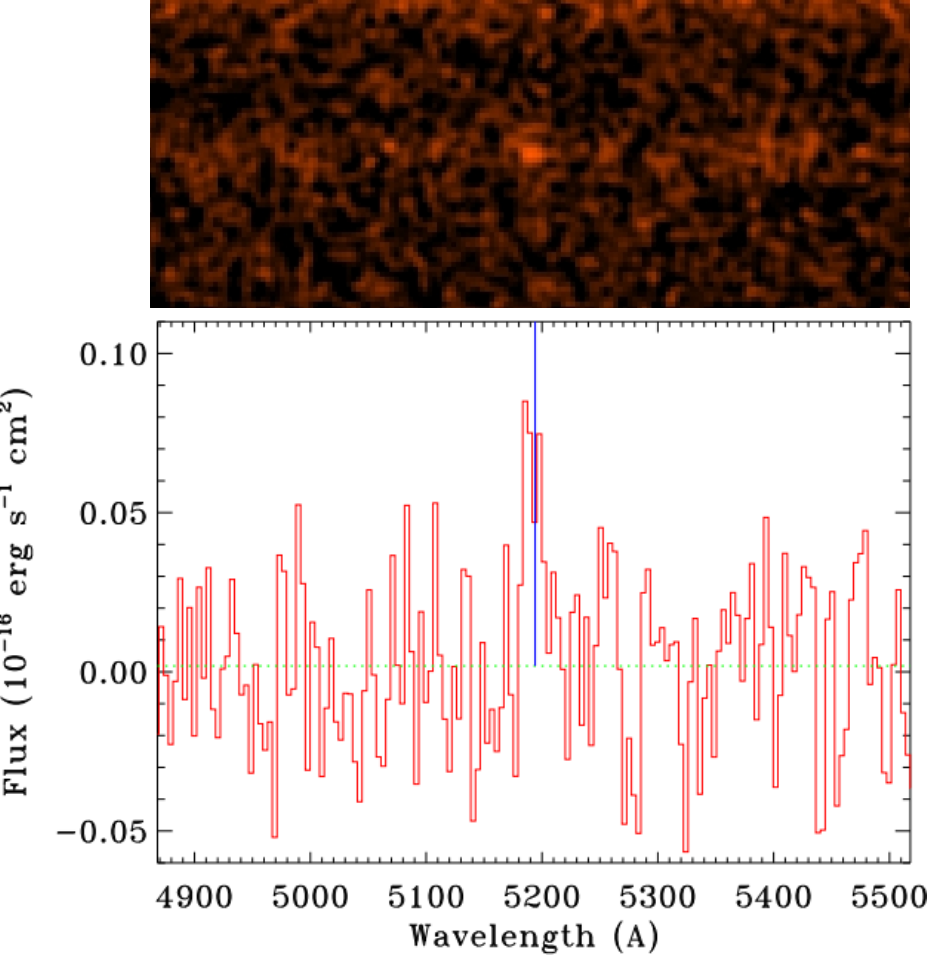}
    \caption{Spectra obtained from the NTT observations for the four
      HzRG candidates in which we detected an emission line, most
      likely \lya, at an S/N$>$5. From left to right, we show
      NVSS~J021439-002359, NVSS~J220526-291029, NVSS~J221708-325449,
      and NVSS~J231647-344233. In the top panels, we show the two
      dimensional spectra, convolved with a Gaussian filter with a
      FWHM of 0\farcs5 for visual purposes (the white area in the
      first panel masks a region corresponding to a bright sky
      line). The spatial extent of the spectra is 10\arcsec. In the
      bottom panels, we present the optimally extracted spectra with
      the same wavelength range as in the top panels. The dashed
      horizontal green lines mark the continuum level, and the vertical
      blue lines mark the center of the \lya\ line from which we
      derived the redshift reported in Table 1.}
    \label{NTT}
\end{figure*}

\begin{table*}[h]
  \caption{Properties of the four confirmed HzRGs.}
    \centering
    \begin{tabular}{l | c r r c | c c c c| c c r }
      Name  & z & F(\lya) & Width & EW & $g$ & $r$ & $i$ & K$_s$ & F$_{76}$ & F$_{1.4}$ & $\alpha_r$ \\
      \hline
NVSS~J021439-002359 & 3.642 & 1.85$\pm$0.30 & 1530$\pm$250  & $>$150      & 26.41 & 24.85 & 24.79 &  ---   & 0.09 & 0.005 & -1.28    \\ 
NVSS~J220526-291029 & 3.448 & 2.48$\pm$0.23 & 1070$\pm$ 90  & $>$290      & 24.71 & 23.35 & 23.13 & 19.99  & 1.39 & 0.137 & -0.80    \\ 
NVSS~J221708-325449 & 3.799 & 2.56$\pm$0.30 &  900$\pm$ 80  & $>$130      & 25.27 & 23.85 & 23.87 & 21.45  & 0.82 & 0.064 & $<-1.3$  \\ 
NVSS~J231647-344233 & 3.272 & 0.82$\pm$0.16 & 1190$\pm$190  & 440$\pm$210 & 21.88 & 20.31 & 19.89 & 18.59  & 2.26 & 0.278 & -0.72    \\    
  \hline
    \end{tabular}
    \smallskip
\tablefoot{Properties of the four confirmed HzRGs: NVSS name,
      estimated redshift of the source, \lya\ flux in 10$^{-16}
      \ergscm$ units, line width (in \kms), observed EW (in \AA), and the
      magnitudes in the $g$, $r$, $i$, and K$_s$ bands. We also report
      the radio flux (in Jy) at 76 MHz from the GLEAM catalog (except
      for NVSS~J021439-002359, for which we list the flux density at
      150 MHz), the flux density at 1.4 GHz from the National Radio
      Astronomy Observatory Very Large Array Sky Survey (NVSS;
      \citealt{condon98}), and the radio spectral index $\alpha_r$.}
    \label{tab1}
\end{table*}

The spectra were reduced with the EsoReflex pipeline
\citep{freudling13}, version 2.3.9, and they were flux calibrated
using observations of standard spectrophotometric stars. The resulting
2D spectra were visually inspected by searching for emission lines,
which were found in four sources at an S/N$>$5. In Fig. \ref{NTT} we
present the 2D spectra and the 1D spectra we extracted at the location
of the emission line.  Their observed equivalent width (EW), measured
against the continuum redward of the lines, are all lower limits
except in one case because no significant continuum is detected. The
derived values are all quite high. They range from EW $>$ 130 $\AA$ to
$>$ 290 $\AA$ and lie in the same range of values found for the
\lya\ line in other HzRGs. The full width at half maximum of the lines
(see Table \ref{tab1}) is in the range FWHM $\sim$ 1,000-1,500 \kms
(well resolved given the resolution of the observations of
$\sim$100-150 \kms) and is also similar to those observed in HzRGs
\citep{rottgering97}. Together with the finding that these lines are
isolated, these results suggest to identify them as \lya\ emission.

The observed emission lines are relatively narrow, compatible with an
identification of a radio galaxy and not of a radio QSO as these
latter sources usually have much broader \lya\ line widths (e.g.,
\citealt{vandenberk01}). The \lya\ fluxes were estimated by direct
integration on the spectra within a spectral range of 30 to 40
\AA\ (i.e., 1,500-2,000 \kms), depending on the source. The line center
was estimated as the barycenter of the emission line, and the
corresponding redshifts of the sources are presented in Table
\ref{tab1} together with the line fluxes, line width, and EW.

We tested the possibility of an erroneous line identification by searching
for alternatives. The most likely alternative is the
[O~II]$\lambda\lambda3726,3729$ doublet (the [O~III]$\lambda5007$ and
the H$\alpha$ lines can be excluded because they are not isolated), which would locate these sources at $z\sim 0.4-0.6$. We explored the
spectra in the regions in which the H$\beta$ and [O~III] emission lines
are expected in this redshift range. With the exception of the
source at the highest redshift (NVSS~J221708-325449), where
these lines are not covered by our spectra, we did not find any
emission line at the expected wavelength. This strengthens the
conclusion that this is indeed the \lya\ line. Furthermore, the
measured line widths are uncomfortably large ($\sim 1,000-1,500$ \kms)
for an assumption that they are produced by these forbidden transitions.

For the remaining sources, we detected no emission
  line, and the continuum emission is at most barely
  visible in general as well. This is expected given the magnitude distribution of the
  targets (see Fig. \ref{histograms}) and the observational
  setup. Thus, we cannot derive any information about their
  nature.

\subsection{Comments on the individual sources}

\noindent
$\bullet$ NVSS~J021439-002359 is associated with the TGSS source
J021439.2-002406. Its radio spectral index $\alpha$ (defined with the
spectrum in the form $f_\nu \propto\nu^{\alpha}$) between 76 MHz and
1.4 GHz is $\alpha = -1.28$. It is unresolved in the VLASS images,
implying a source size $<$7 kpc. The optical counterpart is the HSC
source ID 40678382584728383.

\noindent
$\bullet$ NVSS~J220526-291029 is associated with the EGC source
GLEAM~J220526-291026. Its radio spectral index is $\alpha =
-0.80$. The source is unresolved in the VLASS images, again implying a
source size $<$7 kpc. The host galaxy is identified with the optical
source KiDSDR4~J220526.203-291030.12. Its magnitude in the K$_s$ band
agrees within 0.1 magnitudes with the K-z relation found by
\citet{willott03b}. The K-z relation is defined up to $z\sim4$, where
it has a spread of about one magnitude.

\begin{figure*}
  \center
    \includegraphics[width=0.24\textwidth]{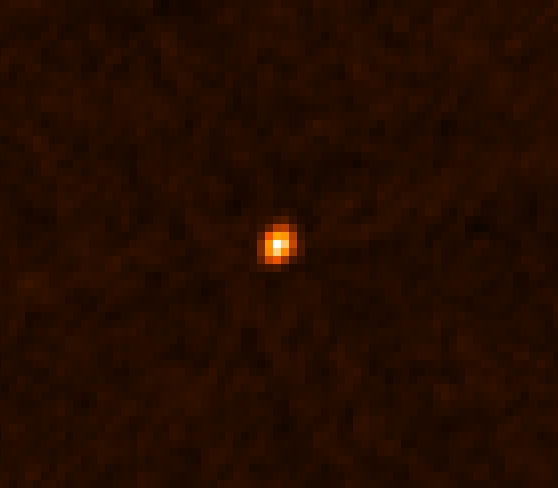}
    \includegraphics[width=0.24\textwidth]{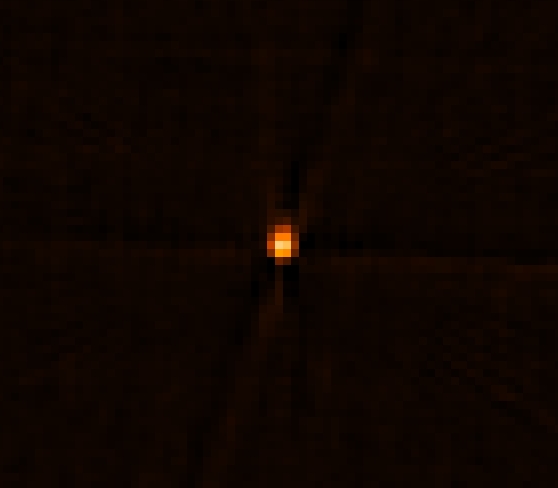}
    \includegraphics[width=0.24\textwidth]{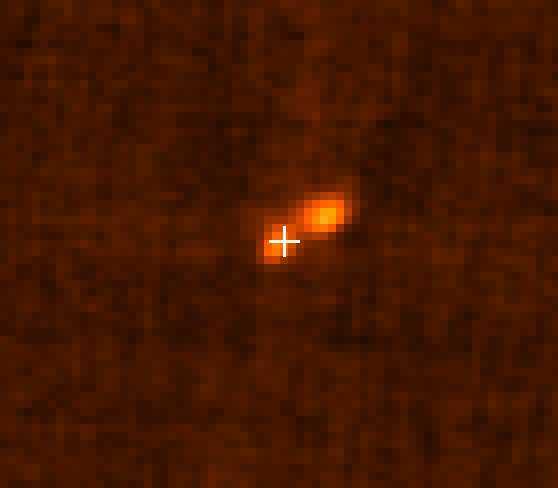}
    \includegraphics[width=0.24\textwidth]{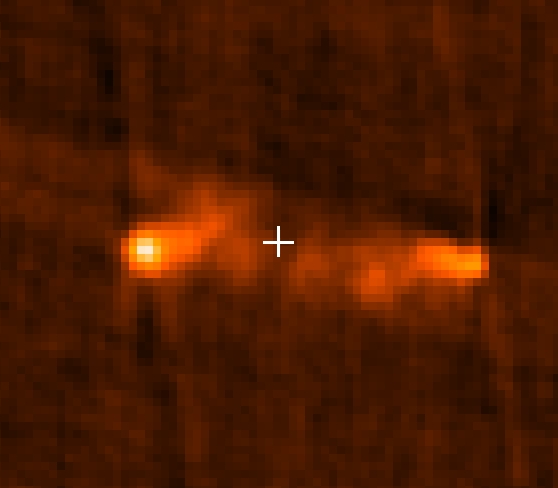} 
    \caption{VLASS images at 3 GHz (field of view
      1$\arcmin\times1\arcmin$) of the four HzRGs showing an emission
      line in their spectra. From left to right: NVSS~J021439-002359,
      NVSS~J220526-291029, NVSS~J221708-325449, and
      NVSS~J231647-344233. The white crosses mark the location of the
      g-dropout optical source in the multiple radio sources.}
    \label{radio}
\end{figure*}

\noindent
$\bullet$ NVSS~J221708-325449 is associated with the EGC source
GLEAM~J221708-325443. In the VLASS images, it shows two components
separated by $\sim 8\arcsec$ (see Fig. \ref{radio}) with flux
densities of 6.3 and 14.9 mJy for the east and west sources,
respectively. The $g$-dropout source (KiDSDR4~J221708.661-325450.93)
is located 0\farcs5 from the eastern radio component. Due to the
presence of two radio sources that are both included in the radio flux density
measurements at low frequencies, we can only estimate an upper limit
to the radio spectral index of the individual sources. By assuming
that the low-frequency flux is entirely associated with the radio source
that is cospatial with the $g$-dropout source, we derived $\alpha > -1.3$. The
host galaxy is underluminous with respect to the K-z relation by
$\sim 1.5$ mag. The \lya\ emission extends over $\sim3\farcs5$ ($\sim$
25 kpc) and shows indications of ordered rotation, with a full amplitude
of $\sim 200$\kms. The extent and velocity field are similar to those of other \lya\ nebulae associated with HzRGs (see.,
e.g., \citealt{wang23}).

\noindent
$\bullet$ NVSS~J231647-344233 is associated with the EGC source
GLEAM~J231647-344228 and is also known as PKS~2314-349. Its radio
spectral index is $\alpha = -0.72$. In the VLASS images, it shows a
large-scale double morphology of $\sim 40\arcsec$ , $\sim 300$
kpc (see Fig. \ref{radio}). The $g$-band dropout optical source
(KiDSDR4~J231647.695-344233.73) is located close to the midpoint
between the two brightest regions in the radio lobes. In addition to
fulfilling the $g$-dropout criteria and having a rather broad
emission line (FWHM = 1190 \kms), this source is also detected in the
$u$ band. The high value of the color $u-g$=1.75 further supports its
identification as a high-redshift source because the Lyman limit at
z=3.272 falls into the $u$ band. It is overluminous with respect to
the K-z relation by about one magnitude.

\section{Efficiency of the spectroscopic program}

This pilot project of a spectroscopic confirmation returned four genuine
HzRGs out of 39 candidates. However, this result is mostly driven by
the poor confirmation rate (only one out of 29 targeted objects) of
g-dropouts that are located at the midpoint of double radio sources. Limited
to sources that are cospatial with a radio component, the success rate
increases to 30\%, with three confirmed HzRGs out of ten attempts.

The selection of HzRG candidates based on the color-color diagrams is
plagued by a large fraction of contaminants. Different classes of
contaminants have been considered in the literature (see, e.g.,
\citealt{davies13,vulcani17}). One of these are cold stars, mainly
brown dwarfs, but this possibility is rejected for the targets that are
cospatial with a bright radio source, except for the unlikely
possibility of a chance alignment. Stars are instead a possible
contaminant for the sources that are found at the midpoint between two radio
components because the direct radio-optical association is lost in this
case. The very different success rate for the two groups of g-dropout
sources, separated based on the morphology of the radio emission,
suggests that this is the dominant effect for those associated with
extended radio sources. In the following, we limit our discussion
to the optical sources that are cospatial with the radio emission, for which we
still need to identify the origin of the substantial fraction (70\%)
of apparently false associations.

In addition to stars, another possible class of contaminants are emission-line galaxies (ELGs), either AGN or star-forming galaxies, in which
the location of lines with a high EW mimics the broadband colors typical
of high-z sources. This is the case, for example, of AGN in which the
[O~III] line falls into the $r$ band, that is, for $0.1 \lesssim z
\lesssim 0.4$. However, if the contaminants were indeed ELGs, the
spectroscopic observations would have revealed them.

The observed drop in the $g$ band, finally, might be due to the 4,000
\AA\ break typical of galaxies dominated by an evolved stellar
population when they are located at $z\sim 0.2-0.5$: the break falls
into the $g$ band and produces the colors required for the
selection. These sources must also be strong radio emitters: Given our
radio flux density threshold, they should have a luminosity of at least
10$^{32}$ erg s$^{-1}$ Hz$^{-1}$, but they must have
emission lines with a low EW to be undetected in their spectra. A
class of objects like these indeed exists, and it is represented by low-excitation
radio galaxies (see, e.g., \citealt{laing94,tadhunter98}).  However,
these low-z radio galaxies are invariably associated with massive
early-type galaxies that span a range of absolute magnitudes from
$M_r \sim -21$ to $M_r \sim -24$ (see, e.g.,
\citealt{buttiglione10,capetti17,baldi18}). When they located at
z=0.5, for instance, even the faintest of these sources would be seen as a $r \sim
20$ source, substantially brighter than the targets observed in our
program (see Fig. \ref{histograms}).

These results cast doubts on the idea that the success rate of our
program is limited by a general misidentification of high-redshift
sources. It suggests that this might instead be due to a flux of their
emission lines that is below our detection threshold.

To plan our observations, we used the well-known correlation between
line and radio luminosity in radio galaxies to estimate the
expected \lya\ fluxes. Low-redshift sources show a strong trend of
an increasing line luminosity with radio power (see, e.g.,
\citealt{baum89a,baum89b,rawlings89,willott99}). More recently,
\citet{buttiglione10} explored this connection based on a
spectroscopic study of the radio sources in the Third Cambridge
Catalog (3C; \citealt{spinrad85}) with z$<$0.3. When they separated the
different spectroscopic classes, \citeauthor{buttiglione10} found that
both low- and high-excitation galaxies follow a quasi-linear
correlation between line and radio luminosity. This implies that the
ratio of the line and radio flux density is approximately
constant. These results were interpreted as evidence for a common
central engine in all powerful radio sources \citep{rawlings91}:
Although radio and line emission are produced by very different
physical processes (synchrotron emission from relativistic electrons
and atomic transitions in gas photoionized by the nuclear light,
respectively), they must be closely linked. \citet{capetti23} found
from observations of 3C radio galaxies at 0.3$<$z$<$0.8 that although
the trend of an increasing line luminosity with radio power is still
present, the slope of radio-line correlation decreases slightly at the
highest radio power, $\sim 10^{35}$ erg s$^{-1}$ cm$^{-2}$ Hz$^{-1}$.

\begin{figure}
  \center \includegraphics[width=0.47\textwidth]{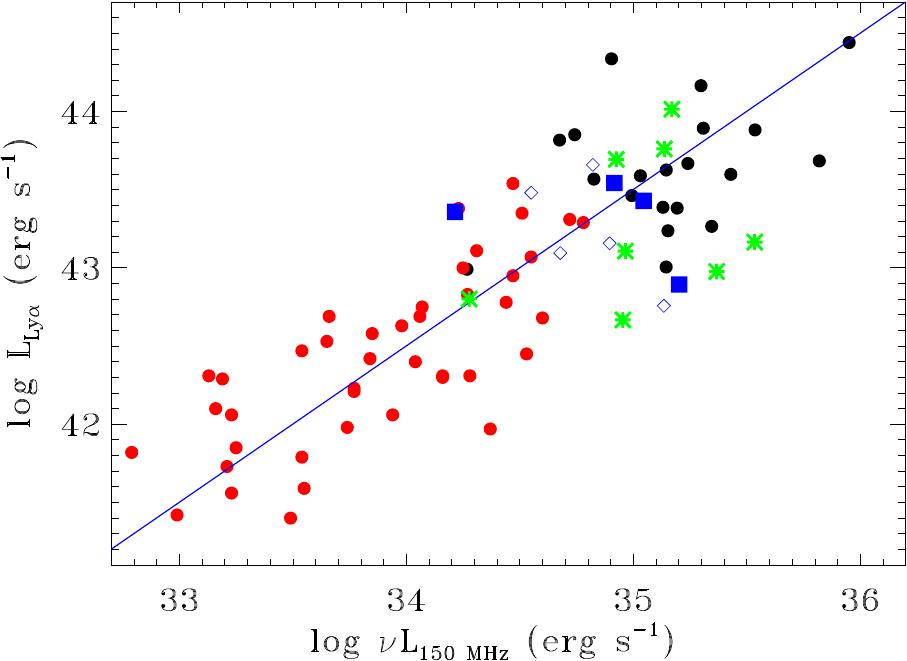}
    \caption{Comparison of the radio and \lya\ luminosities for
      various samples of RGs. The black circles show HzRGs at
      z$\sim$2$-$3.5 from \citet{roettgering97}, the red circles show high-excitation 3C RGs with z$<0.3$ from Buttiglione et al. (2010),
      having assumed Ly$\alpha$/H$\alpha = 5$, the blue diamonds mark
      the five RGs at $z>4$ cited in the text, the
      green stars show eight HzRGs with MUSE observations, and the blue
      squares show the four HzRGs confirmed by this study. The blue
      line represents the locus of a constant ratio of L$_{\rm
        Ly\alpha}$ and L$_{\rm radio}$.}
    \label{lyaradio}
\end{figure}

In order to test whether this behavior is also present in HzRGs, we
collected the measurements of the Ly$\alpha$ fluxes of 19 such sources
at z$\sim2-3.5$ from \citet{rottgering97}. The median ratio for these
HzRGs is $F_{{\rm Ly}\alpha}/F_{150 {\rm MHz}} \sim 5\times 10^{-16}$
erg s$^{-1}$ cm$^{-2}$ Jy$^{-1}$, with an rms of a factor $\sim 3$ (see
Fig. \ref{lyaradio}). This value agrees well with what is
observed at lower redshift, assuming a ratio Ly$\alpha$/H$\alpha$ = 5,
as measured in local AGN \citep{kinney91}. We also considered five
recently discovered HzRGs at $z>4.5$
\citep{jarvis09,saxena18,saxena19,yamashita20}. For these sources, the
line-to-radio ratio is slightly higher, $F_{{\rm Ly}\alpha}/F_{150
  {\rm MHz}} \sim 9\times 10^{-16}$ erg s$^{-1}$ cm$^{-2}$ Jy$^{-1}$.
The addition of the central \lya\ flux measurements of eight HzRGs
with $2.9 < z< 5.2$ for which archival observations with the Multi
Unit Spectroscopic Explorer (MUSE) at the VLT are available (Coloma
Puga, in preparation) confirms the overall trend. These results
  agree with the correlation between \lya\ and radio
  luminosity found by \citet{debreuck00}.

Based on the radio flux densities of our selected HzRG candidates, we
would have expected typical \lya\ fluxes in excess of $\sim 10^{-16}$
erg s$^{-1}$ cm$^{-2}$, which would be easily accessible with one-hour observations
with a 4m class telescope such as the NTT. The four confirmed
HzRGs have indeed \lya\ fluxes in this range.

However, these estimates are based on positive \lya\ detections. These sources might represent the upper end of a
broad distribution of line-to-radio ratios. Most HzRGs might have
much lower \lya\ fluxes at a given radio flux density and remain
undiscovered. The idea that the \lya\ behave differently with
respect to the forbidden lines that are used at low redshift relies on the
properties of the \lya\ emission line. In contrast to the optical
lines, this is a resonant line. Therefore, \lya\ is affected by absorption
effects from both neutral hydrogen and dust, and the emerging
\lya\ emission might be strongly suppressed, depending on the
properties of the intervening medium (see, e.g., \citealt{byrohl20}).

\section{Discussion}

Most of the known HzRGs were selected from samples of USSs, objects in
which the radio spectrum has a steep logarithmic slope. Our results,
combined with those obtained by \citet{jarvis09} and \citet{yamashita20}
(they measured $\alpha = 0.75$ and $\alpha = 0.91 $ for their HzRGs,
respectively) show that when the selection of HzRGs is not based on
the radio properties, but only on the presence of a break in their
rest-frame UV spectra, sources with a wide range of radio spectral
slopes emerge. Therefore, although the search of HzRGs among USS
sources provides a quite high success rate, the resulting objects are
not representative of the whole HzRG population.

Similarly, most known HzRGs are radio sources with a small angular size,
suggesting that this is another general property of these sources.
This can be due to the combination of various effects: the youth of
the sources, the more efficient confinement from the denser external
gas at earlier epochs, or the more effective cooling of the radio-emitting relativistic electrons due to the higher density of the
cosmic microwave background. With our program, we found one HzRG in
which the radio structure extends over $\sim$300 kpc (angular size
$\sim 40\arcsec$). The generally small size of the radio emission in
the known HzRGs is certainly due in part to the difficulty of finding
the host galaxy for sources with a large angular extent. In this case,
a large number of optical sources are possible counterparts,
and it is difficult to isolate the genuine host. This difficulty is
not significantly mitigated by adopting an optical selection based on
the color-color diagrams of HzRG candidates, probably because of the
strong contamination of cold stars with colors typical of
g-dropouts.

When we focused on compact radio sources alone, the selection based on
the color-color diagram we adopted instead yielded a success rate of
$\sim 30$\%. This rate, although based on a small-number statistics,
is substantially higher than that obtained with observations of USSs,
in particular, for the sources at the highest redshift. For example,
\citet{rottgering97} found three radio galaxies with $z>3$ out of 64
USSs for which they were able to determine the redshift. In a
  similar program of USS identification, \citet{debreuck01} confirmed
  10 candidates at z$>$3 out of 62 targets, while \citet{saxena19}
found five such sources (one of which is uncertain) in a sample of 32
candidates, using mostly 8-meter-class telescopes.

Nonetheless, we were unable to find an alternative low-redshift association
to that of genuine Lyman-break galaxies for the seven sources in which
we did not detect emission lines. The failure to detect \lya\ in these
sources suggests that we might have overestimated the expected flux
from this line. We based our estimates on the observed ratio of the
\lya\ and radio flux densities. This estimate is based on
confirmed HzRG candidates, however, and it might correspond to the
upper end of the \lya\ flux distribution. Most of the HzRG population
might still be undiscovered, and the confirmed HzRGs are those with the
brightest line. This possibility relies on the fact that \lya\ is a
resonant line that is strongly affected by absorption effects.

This result can be taken as a suggestion to use different emission lines for the
identification of HzRGs. Rest-frame optical lines are the most obvious
alternative, but the regions of low atmospheric transmission in the
NIR limit their use to reduced ranges of redshift. For
example, the [O~III] line is only visible in sources up to z$\sim$4,
while the [O~II] line falls into the gap between the H and K bands for
sources with $z\sim4-4.5$. 

Active and star-forming galaxies are known to produce bright lines
in the submillimeter band as well, such as those associated with the CO
transitions and the [C~II] line at 157.74 $\mu$m, which are also routinely
observed at high redshift (e.g., \citealt{walter11}).  Two
  HzRGs were indeed confirmed with submillimeter observations
  \citep{drouart20,lee24}. In this case, the main limitation is the
relatively narrow frequency span of a single observation, and multiple
scans might be required to cover the redshift range of interest.

\section{Summary and conclusions}

We presented the results of spectroscopic observations of a sample of
39 candidate HzRGs with the aim to explore the best strategy for building a
statistically sound sample of HzRGs at $z \gtrsim 3$. The targets were selected by combining wide-area radio and optical observations. The candidates
were selected by requiring that their colors fulfilled the criteria of
Lyman-break galaxies. We focused on $g$-dropouts, which are sources that are expected to
be at z$\sim$ 3.0-4.5.

In four of the observed targets, we detected an emission line. Their
large EW and width and the lack of convincing alternatives suggest that they might be
identified as \lya. These confirmed HzRGs are at
z$\sim$3.3-3.8. One of them is located at the center of an extended
radio source with a size of $\sim 300$ kpc, and the others are associated with
compact radio sources. The four newly discovered HzRGs show a wide
range of radio spectral slopes. The search for HzRGs among USS alone
leads to the selection of a subsample of this population that might be biased.

The success rate of our identification of genuine HzRGs with extended
radio structures is very low, only one out 29 attempts. When
we limited the search to those with a compact radio structure, the identification
rate increased to 30\%. This rate is already substantially higher than
the rate obtained with other methods, but we investigated the origin of the
70\% failed confirmations. Bright radio emission
generally excludes a stellar origin, and emission-line
galaxies at low redshift and radio galaxies in which the 4,000
\AA\ break falls into the $g$ band are also excluded.

We concluded that there are no viable alternatives to the
identification of at least part of the HzRG candidates as high-z
Lyman-break galaxies. This suggests that HzRGs have lower \lya\ fluxes
than expected. Our estimate is based on the correlation between
emission line and radio power in known HzRGs. However, they might
represent the upper end of the \lya\ flux distribution. The \lya,
being a resonant line, can display strong absorption and a low escape
fraction.

Other emission lines might be observed to test this hypothesis, such as
forbidden optical lines that are redshifted into the NIR, or submillimeter
emission lines. Both approaches are plagued by the limited spectral
range that can be accessed. The confirmation of HzRG candidates in both cases
would strongly benefit from the possibility of deriving more accurate
photometric redshifts than are currently available.

The forthcoming deep surveys performed by the Rubin Legacy Survey of
Space and Time \citep[Rubin-LSST;][]{ivezic2019} combined with the
Euclid \citep{laureijs10} NIR data will enable us to derive a well-constrained photometric redshift also for the faint host of
HzRGs. This might allow us, on the one hand, to select the candidates
based directly on photo-z, and on the other hand, to search for emission
lines other than \lya. These surveys will also allow us to extend the
quest for HzRGs to a much larger area, and they might paved the way for
the creation of large samples of these elusive sources.

\begin{acknowledgements}
  Based on observations made with ESO Telescopes at the La Silla
  Paranal Observatory under programme IDs 111.24L3.001 and
  114.26ZD.001 and on data products produced by the KiDS
  consortium. The KiDS production team acknowledges support from:
  Deutsche Forschungsgemeinschaft, ERC, NOVA and NWO-M grants; Target;
  the University of Padova, and the University Federico II (Naples).
\end{acknowledgements}

\bibliographystyle{./aa}

\end{document}